# COMPLEXITY

*Carlos Gershenson*
*Instituto de Investigaciones en Matemáticas Aplicadas y en Sistemas,*
*Universidad Nacional Autónoma de México*
*A.P. 20-726, C.P. 01000 México, D.F., México*
*cgg@unam.mx*

The term complexity derives etymologically from the Latin *plexus*, which means interwoven. Intuitively, this implies that something complex is composed by elements that are difficult to separate. This difficulty arises from the relevant interactions that take place between components. This lack of separability is at odds with the classical scientific method—which has been used since the times of Galileo, Newton, Descartes, and Laplace—and has also influenced philosophy and engineering. In recent decades, the scientific study of complexity and complex systems has proposed a paradigm shift in science and philosophy, proposing novel methods that take into account relevant interactions.

## The Limits of Reductionism

Classical science and engineering have used successfully a reductionist methodology, i.e. separate and simplify phenomena in order to predict their future. This approach has been applied in a variety of domains. Nevertheless, in recent decades the limits of reductionism have become evident in phenomena where interactions are relevant. Since reductionism separates, it has to ignore interactions. If interactions are relevant, reductionism is not suitable for studying complex phenomena.

There are plenty of phenomena that are better described from a non-reductionist or 'complex' perspective. For example, insect swarms, flocks of birds, schools of fish, herds of animals, and human crowds exhibit a behavior at the group level that cannot be determined nor predicted from individual behaviors or rules. Each animal makes local decisions depending on the behavior of their neighbors, thus interacting with them. Without interactions, i.e. with reductionism, the collective behavior cannot be described. Through interactions, the group behavior can be well understood. This also applies to cells, brains, markets, cities, ecosystems, biospheres, etc.

In complex systems, having the 'laws' of a system, plus initial and boundary conditions, is not enough to make *a priori* predictions. Since interactions generate novel information that is not present in initial nor boundary conditions, predictability is limited. This is also known as 'computational irreducibility', i.e. there is no shortcut to determine the future state of a system other than actually computing it.

Since classical science and philosophy assume that the world is predictable in principle, and relevant interactions limit predictability, many people have argued that a paradigm shift is required, and several novel proposals have been put forward in recent years.

## The Complexity of Complexity

There is a broad variety of definitions of complexity, depending on the context in which they are used. For example, the complexity of a string of bits, i.e. a sequence of zeroes and ones, can be described in terms of how easy it is to produce or compress that string. In this view, a simple string (e.g, '010101010101') would be easily produced or compressed, as opposed to a more 'random' one (e.g. '011010010000'). However, some people make a distinction between complexity and randomness, placing complexity as a balance between ordered and chaotic dynamics.

A well-accepted measure of complexity is *the amount of information required to describe a phenomenon at a given scale*. In this view, more complex phenomena will require more information to be described at a particular scale than simpler ones. It is important to note that the scale is relevant to determine the amount of information, since e.g. a gas requires much more information to be described at an atomic scale (with all the details of positions and momentums of molecules) than at a human scale (where all the molecular details are averaged to produce temperature, pressure, volume, etc.)

Complexity has been also used to describe phenomena where properties at a higher scale cannot be reduced to properties at a lower scale, i.e. when the whole is more than sum of its parts (see Emergence). For example, a piece of gold has color, conductivity, malleability, and other 'emergent' properties that cannot be reduced to the properties of gold atoms. In other words, there is a potentiality of novel behaviors and properties, i.e. a system with coordinated interacting elements can perform more complex functions than the independent aggregation of the same elements. Emergent properties cannot be reduced to the components of a system because they depend on interactions. Thus, an approach to study complex systems requires the observation of phenomena at multiple scales, without ignoring interactions. Formalisms such as multi-agent systems and network theory have proven to be useful for this purpose.

## Complexity Science?

The scientific study of complexity, under that label, started in the 1980's. Some people argue that it is a science in its infancy, since it has been only a few decades since its inception and it has yet to reveal its full potential. However, some people argue that complexity will never be a science itself, because of its pervasiveness. Since complexity can be described in every phenomenon, e.g. the amount of information to describe it, a science of complexity would be too broad to be useful. A third camp defends that complexity is already a science in its own right. This debate certainly depends on the notion of what a science is. Moreover, one can argue that all three viewpoints are correct to a certain degree. A scientific study of complex phenomena exists; this is not debated. People also agree that this study is offering new insights in all disciplines, and has a great potential, already yielding some fruits. The pervasiveness of complexity is also agreed upon. A scientific approach where interactions are considered, i.e. non-reductionist, has been propagating in all disciplines. Whether people call this approach 'complex' or not, this is not relevant. The ideas and concepts of the scientific study of complex systems are propagating. Perhaps there will never be a science of complexity itself, but complexity is pushing paradigm shifts in all sciences.

## Towards a Philosophy of Complexity

Science has greatly influenced philosophy. For example, Newtonian physics led to philosophical materialism and mechanism. Perhaps unknowingly, the reductionist worldview stemming from classical physics seeped into ontology, where people argued that the only real phenomena were those described by physics, i.e. the laws of matter and energy, while all the rest were only epiphenomena, reducible to Newtonian dynamics in times of Laplace and to elementary particles in recent decades.

Complexity has shown that reductionism is limited, in the sense that emergent properties cannot be reduced. In other words, the properties at a given scale cannot be always described completely in terms of properties at a lower scale. This has led people to debate on the reality of phenomena at different scales. For example, interactions are not necessarily describable in terms of physics, but they can have a causal effect on the physical world. An example can be seen with money, which value is just an agreement between people, i.e. it is not describable in terms of physics. Still, money has a causal effect on matter and energy.

Complexity has also shown the relevance of the observer in the description of phenomena, since depending on the scale at which a phenomenon is described its complexity will change. This is in contrast to classical epistemology, which seeks in objectivism the path to find the 'true' nature of phenomena.

It should also be noted that the novel information generated by interactions in complex systems limits their predictability. Without randomness, complexity implies a particular non-determinism characterized by computational irreducibility. In other words, complex phenomena cannot be known *a priori*.

An interesting feature of the philosophy of complexity is that it is very close to oriental philosophies, which were not influenced by reductionism and consider the relevance of interactions. It could be said that occidental philosophies are rediscovering ideas of oriental philosophies through the scientific study of complexity.

**See also** Emergence, Multi-Agent Modelling, Agent-Based Modeling and Simulation in the Social Sciences, Reductionism in the Social Sciences.

**Further Readings**